\documentclass[twocolumn,prl]{revtex4}
\usepackage{amsmath}
\usepackage{amssymb}
\newcommand{\pa}{\partial}
\newcommand{\bS}{{\boldsymbol{s}}}
\newcommand{\bJ}{{\boldsymbol{J}}}
\newcommand{\bzero}{{\boldsymbol{0}}}

\begin{document}

\title{On physical implementability of reinforcement-learning-driven reasoning}

\author{Eiji Konishi}
\email{konishi.eiji.27c@kyoto-u.jp}
\address{Graduate School of Human and Environmental Studies, Kyoto University, Kyoto 606-8501, Japan}

\date{\today}

\begin{abstract}
In the machine learning community, the reinforcement-learning-driven reasoning process has been intensively studied since around 2025.
In this article, we present a negative result on the hardware implementation of this reinforcement-learning-driven reasoning process in a physical neural network model in the presence of a reward external potential energy for a given reward.
In this hardware implementation, the existence of a chain of thought in dynamics contradicts the Lyapunov stability of the system.
\end{abstract}

\maketitle

Machine learning research has recently advanced as a result of a shift in fundamental direction from conventional approaches to generative artificial intelligence (e.g., generative adversarial networks, diffusion models, and transformers and the like) to reasoning machine learning accompanied by a chain of thought (CoT) \cite{DL1,RL,DL2,CoT,DPT,DeepSeek}.
This shift is based on Kahneman's double-process theory \cite{Kahneman}, where system 1 and system 2 in this theory correspond to the fast and the slow processes of thinking, respectively \cite{DPT}.
In particular, system 2 (slow process) is considered to play a fundamental role in future artificial general intelligence (AGI).
In 2025, DeepSeek investigated reinforcement-learning (RL)-driven self-correction by adopting group relative policy optimization, instead of proximal policy optimization, as online RL \cite{DeepSeek}.
The aforementioned investigation was implemented in software, that is, machine learning.
In this article, we examine whether reasoning via RL-driven self-correction can be physically implemented, where physical convergence to a {\it given} answer state as the ground state of the model is guaranteed.
Namely, we examine whether there exists a hardware counterpart of Ref. \cite{DeepSeek} for a reduced purpose (i.e., a physical derivation of a given answer state).

As a concrete model of the hardware implementation of both system 1 and system 2, a binary-state classical physical neural network \cite{Hopfield,Rojas,Comment}\footnote{For the extension of this network to incorporate projective quantum measurement as both quantum and classical measurements, see Ref. \cite{Konishi1}.} is considered, where we drop the condition of the symmetric synaptic couplings from the Hopfield network.
This neural network, which is a parallel physical computer, could potentially be integrated into the hardware of a von Neumann-type series computer, which verifies, intervenes, and serializes online RL, as its nested series-parallel hybrid extension to perform rapidly and physically converging reasoning via RL-driven self-correction.
Here, we model the process of system 2 such that it accompanies the process of system 1, and both system 1 and system 2 correspond to a fast process.

This {\it physical} neural network has the strengths of the synaptic junctions as its slowly tunable logical constants, which are analogous to the physical constants.
In this hardware unit, the topology of the loss function for online RL is mapped into the slowly tunable energy landscape.

Now, we hypothesize the energy function of this physical neural network.
In this energy function, the logical constants have a slowly tunable skew-symmetric part (logical directionality required for reasoning \cite{SK}) in addition to a fixed symmetric part in the presence of the slowly tunable and additively separable (see Eq. (\ref{eq:Phi})) reward external potential energy for online RL for a given reward, which is the answer state to be logically {\it derived}.
The fixed symmetric part is the opposite of the off-diagonal part of the Hessian of the energy function and is obtained by hardware meta-learning; it is typically negative semi-definite for spatially locally stable reasoning \cite{Izhikevich}.
Here, the tuning is the learning.

We assume the following energy function for the neural network:
\begin{equation}
E=-\frac{1}{2}\sum_{i\neq j}J_{ij}s_i s_j +\sum_i \theta_i s_i+\Phi_R(\{s_i\})\;.\label{eq:E}
\end{equation}
Here, $s_i=0,1$ is the binary state of neuron $i$, $\theta_i$ is the bare threshold value for the activation of neuron $i$, and the strength of synaptic junction $J_{ij}$ from neuron $i$ to neuron $j(\neq i)$ is the sum $J_{ij}^S+J_{ij}^A$ of its symmetric part $J_{ij}^S(=J_{ji}^S)$ and skew-symmetric part $J_{ij}^A(=-J_{ji}^A)$.
In this expression, $\Phi_R(\{s_i\})$ is the reward external potential energy on the discrete hypercube (the Hamming space).
Intuitively, whereas the sum of the first and second terms is relatively shallow and spatially local and complex, the third term is relatively deep (dominant energy scale) and spatially global and smooth.

In our hypothetical picture of the dynamical process, by adding the reward external potential energy, the state of the system would spiral down this reward potential in higher dimensions.
The horizontal part in this spiral down represents the CoT driven by the reward potential $\Phi_R(\{s_i\})$ along the logical directionality generated by the skew-symmetric part of $J_{ij}$.
This CoT process is schematically shown as the following sequence of conditional logical transitions:
\begin{equation}
{\rm CoT}:\bS\overset{\bJ^A}{\longrightarrow} \bS\overset{\bJ^A}{\longrightarrow} \bS\cdots\;,
\end{equation}
where $\bS$ and $\bJ^A$ represent a vector of bits (a logical pattern as an attractor) and a skew-symmetric matrix (logical directionality), respectively.

The reward potential energy is an additively separable function with respect to the neurons:
\begin{equation}
\Phi_R(\{s_i\})=\sum_i \phi_i(s_i)\;.\label{eq:Phi}
\end{equation}
This external potential energy property is required so as not to change the logical structure set by the symmetric part (a symmetric quadratic form with no diagonal terms) of the energy function (\ref{eq:E}).

Note that the skew-symmetric part vanishes in the energy function (\ref{eq:E}) but drives a non-conservative flow (vortex) in the discrete-time evolution law of the gradient system \cite{GM}.

In the following, we show a negative result, which asserts that this flow cannot be confined by the reward potential to restore Lyapunov stability at the convergence stage\footnote{Here, the value of the Lyapunov function (energy function) always decreases when the state of the system changes according to the gradient flow \cite{Hopfield,Rojas}.}.
Consequently, our hypothesis on the existence of the energy function (\ref{eq:E}) is false.

In the discrete-time evolution law of the gradient system, the force that acts on neuron $i$ is
\begin{equation}
F_i=-\frac{\pa E}{\pa s_i}+F_i^A\;.
\end{equation}
Here, $F_i$ is the sum of the force of the symmetric part $F_i^S$, the force of the skew-symmetric part $F_i^A$, and the reward force $F_i^R$.
The force ${\cal F}_i$ acting on neuron $i$ additional to that of the Hopfield network, $F_i^S$, is
\begin{eqnarray}
{\cal F}_i&\equiv&F_i^A+F_i^R\\
&=&\sum_{j(\neq i)} J_{ij}^A(s_j-s_j^\ast)-\frac{d\phi_i(s_i)}{ds_i}
\end{eqnarray}
for the answer state $s_j^\ast$ of neuron $j$.

For the function (\ref{eq:E}) to be a Lyapunov function \cite{Rojas}, the reward potential $\Phi_R$ must be tuned by online RL to satisfy the following two conditions for the confinement of the vortex, which is generated by the skew-symmetric $J_{ij}^A$, by the reward potential $\Phi_R$:
\begin{eqnarray}
|F^R_i|&>&|F^A_i|\;,\label{eq:C1}\\
{\rm sgn}(F^R_i)&\neq&{\rm sgn}(F^A_i)\label{eq:C2}
\end{eqnarray}
for all neurons $i$ with an arbitrary state $\{s_j\}$ except for the answer state $s_i^\ast$.
For neuron $i$, these two conditions together equal ${\rm sgn}({\cal F}_i)={\rm sgn}(F^R_i)\neq {\rm sgn}(F^A_i)$.

Next, in the case where the answer state is $\bS^\ast=\bzero$, we consider the additional force ${\cal F}_1$ to $F^S_1$ acting on neuron $1$ in the neural states $s_i=0$ for $i\neq 1,2$.
The additional force ${\cal F}_1$ is
\begin{equation}
{\cal F}_1=J_{12}^As_2 -\frac{d\phi_1(s_1)}{ds_1}\;.
\end{equation}
Then, at $s_1=0$ and $s_2=1$, the conditions (\ref{eq:C1}) and (\ref{eq:C2}) extended to $s_1^\ast$ on the functions $F^R_1$ and $F^A_1$ lead to $|J_{12}^A|$ being equal to $|d\phi_1(0)/ds_1|=0$.
This argument results in $J_{1i}^A=0$ for all $i\neq 1$.\footnote{Without the property (\ref{eq:Phi}), this result would be weakened to $\sum_{i(\neq 1)}J_{1i}^As_i^\ast=0$, which holds identically since $\bS^\ast=\bzero$.}
This result for neuron $1$ holds for all neurons.
This means that there is no CoT (generated by $\bJ^A$) that is compatible with the Lyapunov stability.
This shows our negative result.

This negative result suggests that RL-driven reasoning (in our sense, a derivation of a given reward) needs to be implemented as software (here, machine learning \cite{DeepSeek}) in a von Neumann-type computer, rather than in hardware (in a physical neural network in the presence of a reward potential), which is ruled out by our result.

According to Ref. \cite{Konishi2}, the von Neumann-type computer can be extended to a physically grounded induction machine or agent by incorporating projective quantum measurement as a classical measurement; its evolution is on the route toward the computable models of AIXI \cite{Hutter} as an agent because, in the von Neumann-type computer, the software and hardware are separated and the software is stored as data in the main memory.
On the other hand, the evolution of the physically grounded neural network \cite{Konishi1} is on the route toward the human brain because the software and hardware are inseparable.

The present result suggests that, in the future implementation of AGI as a network between software-sharing series-parallel hybrid computers of the above two types, the individual node is logically governed by the von Neumann-type computer (system 2), and the integrated physical neural network (system 1) with sensor(s) provides only associative memory (rapid convergence without iterative exploration) and projective quantum measurement \cite{Konishi1}.

\end{document}